# Multi-State, Ultra-thin, BEOL-Compatible AlScN Ferroelectric Diodes

Kwan-Ho Kim,[1,†] Zirun Han,[1,2,†] Yinuo Zhang,[1,3] Pariasadat Musavigharavi,[3] Jeffrey Zheng,[3] Dhiren K. Pradhan,[1] Eric A. Stach,[3] Roy H. Olsson III,[1] and Deep Jariwala[1,*]

[1]Department of Electrical and Systems Engineering, University of Pennsylvania, Philadelphia, Pennsylvania 19104, United States

[2]Department of Physics and Astronomy, University of Pennsylvania, Philadelphia, Pennsylvania 19104, USA

[3]Department of Materials Science and Engineering, University of Pennsylvania, Philadelphia, PA, USA

[†]These authors equally contributed to this work.

[*]Author to whom correspondence should be addressed: dmj@seas.upenn.edu

**ABSTRACT**

The growth in data generation necessitates efficient data processing technologies to address the von Neumann bottleneck in conventional computer architecture. Memory-driven computing, which integrates non-volatile memory (NVM) devices in a 3D stack, is gaining attention, with CMOS back-end-of-line (BEOL) compatible ferroelectric (FE) diodes being ideal due to their two-terminal design and inherently selector-free nature, facilitating high-density crossbar arrays. Here, we demonstrate BEOL-compatible, high-performance FE-diodes scaled to 5, 10, and 20 nm FE $Al_{0.72}Sc_{0.28}N/Al_{0.64}Sc_{0.36}N$ films. Through interlayer (IL) engineering, we show substantial improvements in the ON/OFF ratios (>166 times) and rectification ratios (>176 times) in these scaled devices. The superlative characteristics also enables 5-bit multi-state operation with a stable retention. We also experimentally and theoretically demonstrate the counterintuitive result that the inclusion of an IL can lead to a decrease in the ferroelectric switching voltage of the device. An in-depth analysis into the device transport mechanisms is performed, and our compact model aligns seamlessly with the experimental results. Our results suggest the possibility of using scaled $Al_xSc_{1-x}N$ FE-diodes for high performance, low-power, embedded NVM.

**INTRODUCTION**

The expansion in data generation via a range of ubiquitously connected devices has intensified the need for highly efficient and rapid data processing technologies that can overcome the von Neumann bottleneck.[1] Recent research suggests the use of memory-driven computing by integrating emerging non-volatile memory (NVM) devices with the processing transistors in a dense monolithic 3D stack. For this application, a CMOS BEOL-compatible FE-diode is particularly attractive, as it is a two-terminal device with rectifying *I-V* characteristics, enabling selector-free and high-density crossbar array memory technology[2].

The ferroelectric material $Al_{1-x}Sc_xN$ (AlScN) is promising for application in FE-diodes due to its unique ferroelectric properties, such as a high remnant polarization ($P_r$) of > 125 μC/cm$^2$, coercive field ($E_c$) of 3 – 6 MV/cm, and square-shaped polarization-electric field (P-E) loop.[3, 4] Moreover, $Al_{1-x}Sc_xN$ (x < 0.43) is highly CMOS BEOL-compatible with a low process temperature (< 400 °C), and has a single, stable wurtzite ferroelectric phase, unlike $HfO_x$ polymorphisms exhibiting ferroelectricity only in metastable phases.[1] As a result, ferroelectric AlScN can offer uniform and reliable device performance without any additional post-annealing or other phase stabilization procedures.

Despite the remarkable attributes of FE AlScN such as large $P_r$ and moderate ε, scaling of FE-diodes to low-voltages concurrently with high rectification and ON/OFF ratios remains unachieved. The majority of existing studies predominantly focus on the ferroelectric tunneling junction (FTJ), which shows a marginal rectification ratio, necessitating the addition of a selector.[1, 2] While some recent demonstrations have shown AlScN FE-diodes, the thickness of the AlScN films used typically range over 20 - 45 nm.[2, 5] To lower the operating voltage of an AlScN FE-diode, a reduction in AlScN thickness is essential. Encouragingly, recent research have shown that AlScN can be scaled down to sub-10 nm dimensions without compromising its outstanding ferroelectric properties.[6] However, the ON/OFF current ratios and rectification characteristics of the scaled devices are far from desirable values.

One approach to enhancing the performance of FE-diodes is the introduction of an interlayer (IL) in the device stack, resulting in a metal-ferroelectric-insulator-metal (MFIM) structure. The inclusion of an IL can enhance the FE polarization-induced electrostatic modulation of the tunneling barrier in FE-diodes, thereby improving ON/OFF and rectification characteristics.[7] Therefore, IL engineering becomes a key aspect of AlScN FE-diode optimization at these scaled thicknesses to improve performance, which we have thoroughly explored.

Here, we present the successful demonstration of BEOL-compatible FE-diodes using $Al_{0.72}Sc_{0.28}N$ or $Al_{0.64}Sc_{0.36}N$ films with thicknesses of 5 nm, 10 nm, and 20 nm. With the inclusion of an $AlO_x$ IL, our scaled $Al_{0.72}Sc_{0.28}N/Al_{0.64}Sc_{0.36}N$ FE-diodes show a substantial enhancement in both the ON/OFF and rectification ratios. Notably, the 10 nm $Al_{0.72}Sc_{0.28}N$ FE-diode exhibits a remarkable increase in the ON/OFF ratio from 403 to 3173 and an improvement in the rectification ratio from 2992 to 5696. For the 5 nm $Al_{0.72}Sc_{0.28}N$ FE-diodes, these values increase from 5 to 832 and 14 to 2467 for the ON/OFF and rectification ratios respectively. In addition to these performance enhancements, we also present a counterintuitive benefit of IL integration to be a reduction in ferroelectric switching voltage. We verify this effect through two experimental methods of switching voltage extraction as well as theoretical electrostatics analysis. Through our analysis, we show that ferroelectric systems exhibiting high $P_r$, low $\varepsilon$, and square-like hysteresis,[4] such as AlScN, are ideal for exploiting this effect. The use of thicker ILs (>3 nm) also stabilizes intermediate $P_r$ states, allowing for 32 stable memory states with a maximum ON/OFF ratio separation of 1175. Each of these states show robust retention for up to 300 seconds, with two states among these demonstrating remarkably stable retention beyond 50,000 seconds, maintaining an ON/OFF ratio exceeding 700. Finally, we provide a comprehensive analysis of the FE-diode operation, corroborated by a well-fitting compact device model that is based on thermionic emission and the Poole-Frenkel effect.

**RESULTS AND DISCUSSION**

**General Structure of the AlScN FE-diodes**

The FE-diodes in this work consists of 50 nm Al bottom electrodes grown on a sapphire wafer. The ferroelectric layer consists of sputter-deposited 5, 10, or 20 nm $Al_{0.72}Sc_{0.28}N/Al_{0.64}Sc_{0.36}N$. The IL consists of ALD-deposited AlO$_x$ with thickness ranging from 0 (no IL) to 5 nm. The top electrode arrays consist of 20 μm-radius pads of Ti or Cr metal with an Au capping layer. The detailed information of the fabrication processes is described in method section. TEM images of our devices are shown in Figure 1(a) and (b). A detailed compositional analysis is presented in Figure S1 in supplementary information. We performed the TEM on our thinnest AlScN FE-diode devices and the AlScN thickness is confirmed to be 5 nm in this case. The schematic of the fabricated FE-diode is shown in Figure 1(c), and an optical image of the device arrays is shown in Figure 1(d).

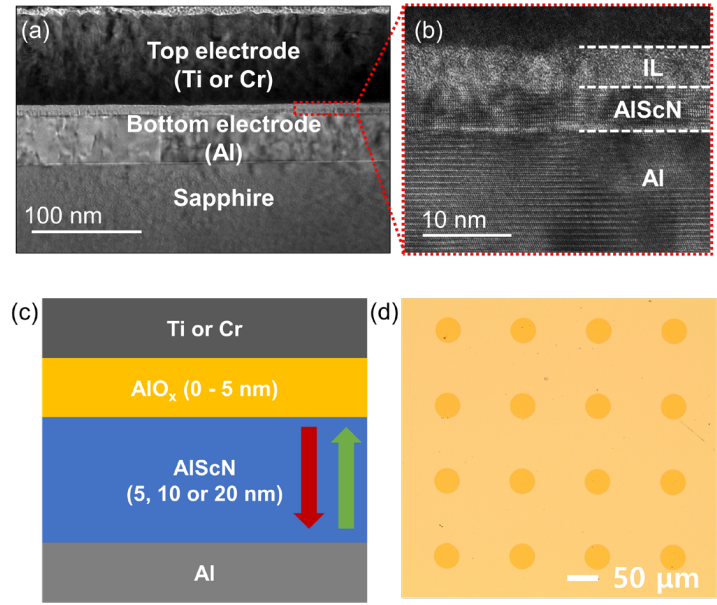

**Figure 1. General structure of the Al$_{0.72}$Sc$_{0.28}$N/Al$_{0.64}$Sc$_{0.36}$N ferroelectric diode devices.** **(a)** Cross-sectional bright field STEM and **(b)** high-resolution TEM image of the device. **(c)** Schematic diagram of the MFIM Al$_{0.72}$Sc$_{0.28}$N/Al$_{0.64}$Sc$_{0.36}$N ferroelectric diode. **(d)** Optical image of the device array. The darker contrast circles represent the top metal electrodes of 20 μm radii.

## I-V Characterization of 20 nm AlScN FE-diodes

We start by describing our results on FE-diodes built on 20 nm thick Al$_{0.72}$Sc$_{0.28}$N and later make comparisons with thinner devices. Figure 2(a-f) presents the experimental *I-V* characteristics of FE-diodes that consist of a 20 nm $Al_{0.72}Sc_{0.28}N$ layer, an $AlO_x$ IL with thicknesses varying from 0 to 5 nm, an Al bottom electrode, and a Cr or Ti top electrode. The driving voltage is applied to the bottom electrode while the top electrode is grounded. The devices display an ON/OFF ratio and asymmetric *I-V* characteristics between the positive and negative side in devices with both Cr (Supplementary Information Figure S7) and Ti (Figure 2) top electrodes. The inclusion of Cr allows us to verify that the resistive switching is triggered by the ferroelectric switching and is not due to Ti atom diffusion forming a current path as is the case in conductive bridge oxide memristors[8].

The ON/OFF and rectification ratios of the 20 nm $Al_{0.72}Sc_{0.28}N$ FE-diodes are extracted from the *I-V* characteristics and plotted against IL thickness in Figure 2(g-h). We extracted the rectification ratio of the devices at the point of the maximum ON/OFF ratio. This is because it is preferable to apply the read voltage at the point of largest ON/OFF ratio, and a large rectification ratio at this point enhances self-rectification characteristics that suppress sneak current paths in crossbar arrays. We observed a significant increase in both ON/OFF ratio and rectification ratio vis-à-vis control samples at large (>3 nm) IL thicknesses. To interpret these results, we develop a precise physical model for the *I-V* characteristics, which is described in the following section.

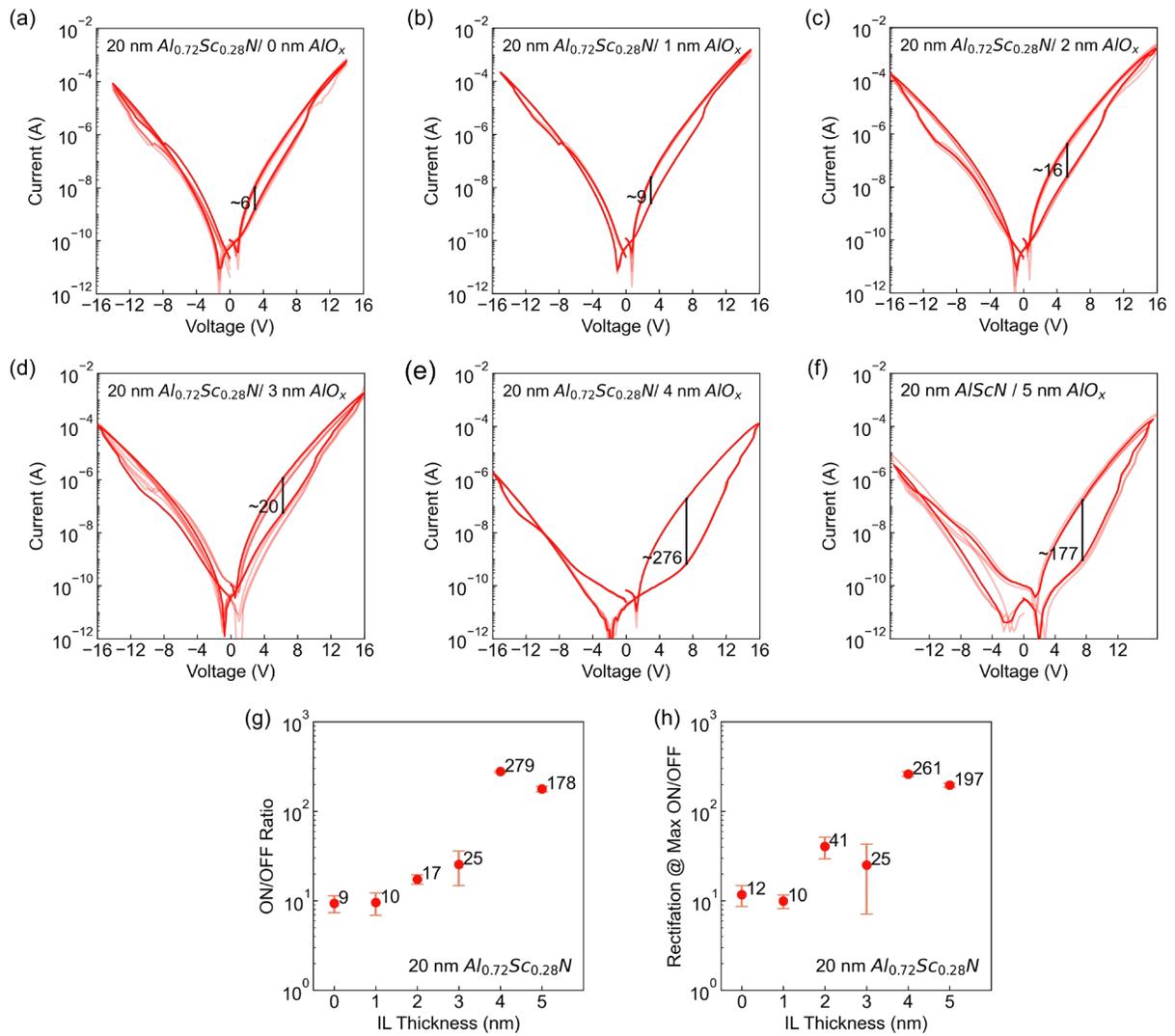

**Figure. 2. I-V curves and extracted performance characteristics of 20 nm Al$_{0.72}$Sc$_{0.28}$N ferroelectric diodes.** I-V curves of 20 nm Al$_{0.72}$Sc$_{0.28}$N ferroelectric diodes with an AlO$_x$ thickness of **(a)** 0 nm, **(b)** 1 nm, **(c)** 2 nm, **(d)** 3 nm, **(e)** 4 nm, and **(f)** 5 nm are shown. For all devices, Al is the bottom electrode, and Ti is the top electrode. For the statistical analysis, each I-V curve is obtained from five different devices. **(g)** Extracted IL-thickness dependent ON/OFF ratio of the devices. **(h)** IL-thickness dependent rectification ratio extracted at the point of the maximum ON/OFF ratio.

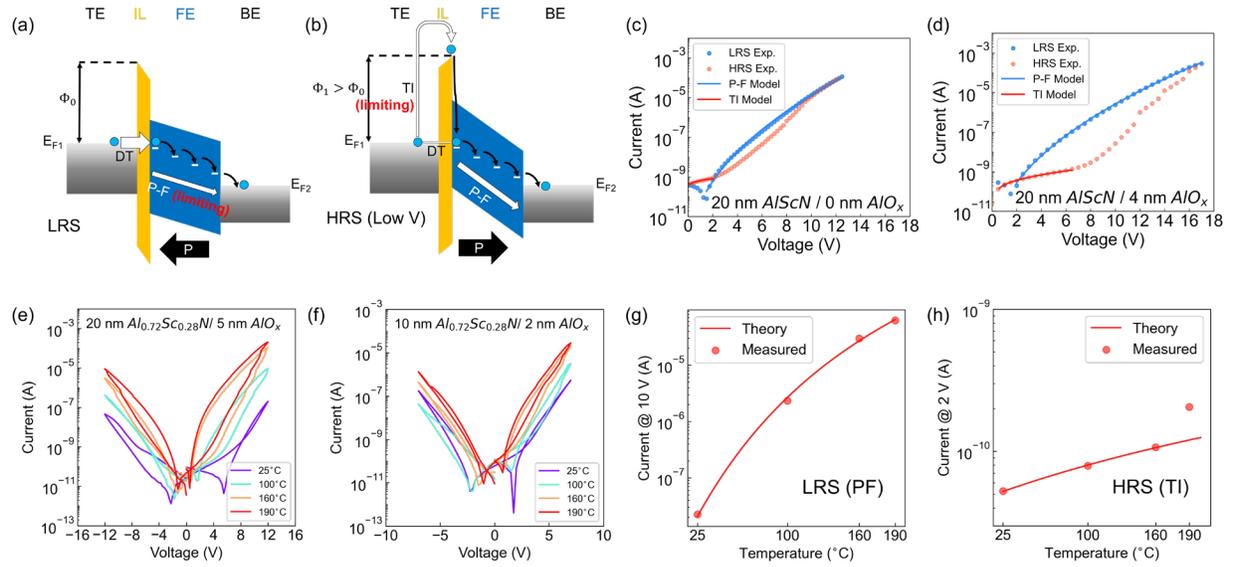

**Figure 3. Theoretical analysis of carrier transport through the device and compact model fitting.** Energy band diagram and active transport mechanisms of the ferroelectric diode for **(a)** LRS **(b)** HRS. The diagrams show the relative importance of various tunneling mechanisms. They also show the modulation of local electric fields and the emission barrier Φ by the polarization state of the ferroelectric. **(c)** Fitting of the Poole-Frenkel (P-F) and thermionic emission (TI) models to the 20 nm device with no IL. **(d)** Fitting of the two carrier transport models to the 20 nm device with 4 nm $AlO_x$ IL. This demonstrates the extension of the active range of TI that induces an enhanced ON/OFF ratio. The temperature dependent *I-V* curves of (e) 20 nm $Al_{0.72}Sc_{0.28}N$/5 nm $AlO_X$ device, and (f) 10 nm $Al_{0.72}Sc_{0.28}N$/2 nm $AlO_X$ device. (g) The extracted current values from *I-V* curve of 20 nm $Al_{0.72}Sc_{0.28}N$/5 nm $AlO_X$ device according to temperature for in LRS state at 10 V, and (h) HRS state at 2 V.

## Compact Modeling of AlScN FE-diode I-V Characteristics

To understand the *I-V* characteristics of our FE-diodes in detail we have developed a 1D compact model accounting for all electrostatics as well as band and defect induced transport within the band-gap. Due to neutral-level traps in AlScN created by nitrogen vacancies[9] and the polarization-induced lowering of the tunneling barrier to these traps as illustrated in Figure 3(a), the LRS *I-V* characteristics of our devices are well-described by the bulk-limited Poole-Frenkel (P-F) emission model as shown in the blue curves in Figure 3(c, d).

In contrast, the HRS characteristics contain several distinct regions. Of particular interest is the low-bias, low-current regime that supports the largest ON/OFF ratio observed in the device.

In this regime, the tunneling barrier to traps in AlScN is large. As such, the current is interface-limited, where direct tunneling (DT) to trap sites[10] and thermionic emission (TI) over the barrier[11] are two possible mechanisms for carrier injection into AlScN (Figure 3(b)). The interplay between these current injection mechanisms is key to understanding the enhancement in our device performance. As IL thickness is increased, the capacity for carrier injection via TI stays roughly constant as it is an interfacial mechanism. However, we gain an increased ability to modulate and suppress carrier transport via DT using the polarization state of the ferroelectric, which we will proceed to analyze quantitatively.

An important feature of our IL-inserted devices is that the local electric field in the IL can be significantly modulated by the polarization state of the ferroelectric, and the resulting voltage drop across the IL due to $P_{FE}$ can be written as

$$\Delta V_{IL} = E_{IL} t_{IL} = \frac{P_{FE} t_{IL}}{\epsilon_{IL} + \epsilon_{FE} \frac{t_{IL}}{t_{FE}}} \tag{1}$$

This follows from equation ES2 and ES3 in the SI. As such, the maximum electrostatic potential difference in the IL between the HRS and LRS would be $2\Delta V_{IL}$ (attained at the IL-ferroelectric interface) while the average potential difference between the two states in the IL is $\Delta V_{IL}$.

As the IL thickness increases, the amount of barrier modulation $\Delta V_{IL}$ also increases. For example, for our 20nm AlScN / 4 nm AlO$_x$ devices, $\Delta V_{IL} = 3.7V$ (equivalently an electric field modulation by 9.25 MV/cm) even if we assume a moderate $P_{FE}$ of 10 µC/cm$^2$ due to depolarization effects. This has a significant impact on tunneling transport through the IL. In the LRS, the tunneling barrier would be nearly triangular, so carrier transport across the IL approaches the Fowler-Nordheim regime where the effective tunneling distance for carriers is shorter than the IL thickness. This mitigates the suppression of tunneling current by the increase in $t_{IL}$. On the other hand, in the HRS, carriers must tunnel through the entire thickness of the IL with an increased barrier height. This means that even at large (3-5 nm) IL thicknesses, in the LRS, the IL tunneling barrier will still be nearly transparent in comparison to the PF transport across the AlScN. On the other hand, it grows large enough in the HRS so that the direct tunneling contribution to the current is negligible compared to TI at low voltages. As such, the increased control of DT current via polarization-dependent barrier modulation explains the enhancement of the ON/OFF ratio we observe as IL thickness increases. This also means that for low biases or thick barriers, we can take TI to be the limiting mechanism for HRS carrier transport as illustrated by the red curves in Figure 3(b).[11]

To further confirm the validity of our device model, temperature-dependent electrical measurements are performed as shown in Figure 3(e, f). We performed the measurements by heating our 20 nm $Al_{0.72}Sc_{0.28}N$/5 nm $AlO_x$ FE-diodes using a MHP30 hotplate to 100, 160, and 190 °C. The device temperatures are verified using a digital multimeter equipped with a K-type thermocouple. At these temperatures, we performed standard DC *I-V* measurements using a Keithley 4200A-SCS. These I-V curves (including at room temperature) were obtained using a lower operating voltage range compared to our other measurements at room temperature because the $E_C$ of AlScN decreases with temperature. As such, applying the maximum voltage required for full switching at room temperature at high temperature would cause the devices to fail. The data obtained from these measurements was fitted against the PF and TI transport models and an excellent agreement was confirmed as shown in Figure 3(g, h). In particular, PF predicts the scaling of current density with respect to temperature as $J \propto e^{-1/T}$ at a fixed voltage while TI predicts that $J \propto T^2 e^{-1/T}$. These trends are analyzed and again an excellent agreement was shown, as can be seen in Figure 3(g, h). In the HRS TI fitting (Figure 3(h)), the points obtained at 190 °C deviates from the trend line, and it is speculated that the device operation at 190 °C is unstable. The temperature-dependent *I-V* curves are shown in Figure S8 in the Supplementary Information. These temperature dependent I-V curves solidify that the transport mechanisms described in the manuscript are valid descriptions for our devices.

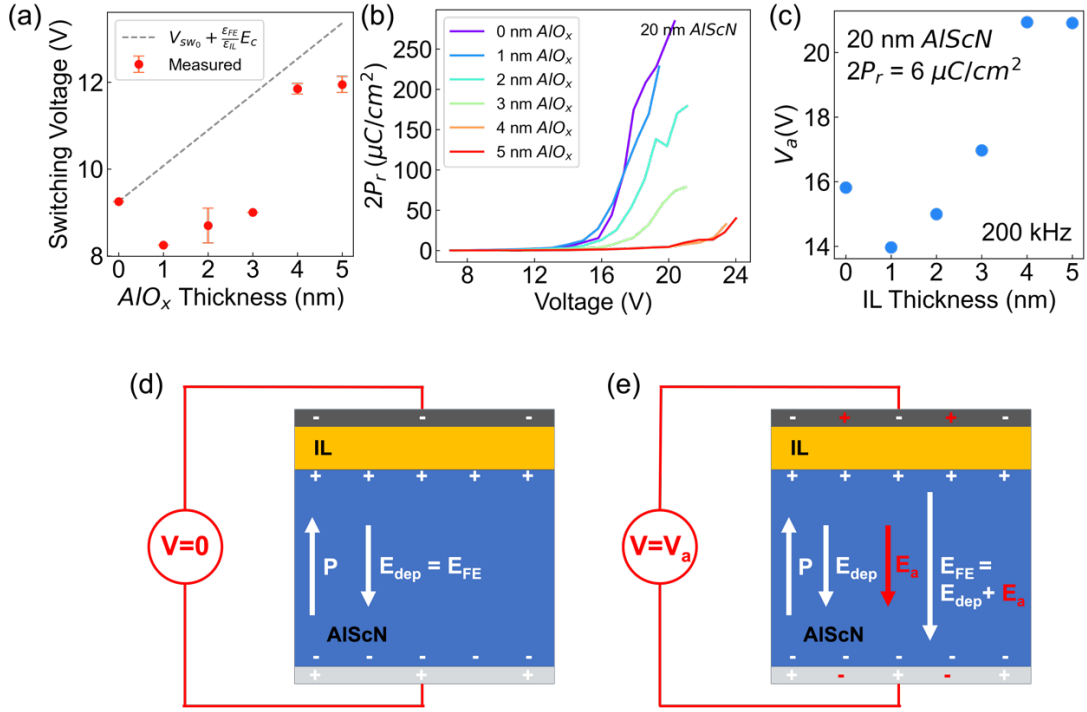

**Figure 4. IL-induced switching voltage reduction and electrostatic analysis of the device structure. (a)** Ferroelectric switching voltages of 20 nm $Al_{0.72}Sc_{0.28}N$ FE-diodes extracted from quasi-DC *I-V* curves as shown in figure 2. The switching voltages are extracted at the first abrupt peaks of the derivatives of the log *I-V* curves. The "standard expectation" of $V_{sw}$ scaling with IL thickness without consideration of the depolarization effect is shown with a gray dotted line. **(b)** PUND measurement results using 200 kHz pulses for the 20 nm device with $AlO_x$ thicknesses from 1 to 5 nm. Voltage is applied in 1 V steps until the point of device breakdown. **(c)** Applied voltage ($V_a$) extracted from the PUND measurements needed to reach a $2P_r$ value of 6 μC/cm² for various IL thicknesses. **(d)** Graphical representation of electrostatics in the FE-diode when external voltage is 0 V, and **(e)** $V_a$.

**IL-induced Switching Voltage Reduction in AlScN FE-diodes**

We make an interesting and counterintuitive observation in our AlScN FE-diodes. Notably, the ferroelectric switching voltage ($V_{sw}$) of our FE-diodes can decrease with increasing IL thickness. This is counterintuitive since any non-ferroelectric insulating layer in a FE capacitor stack will cause a voltage drop across it, which would decrease the electric field in the ferroelectric. The electric field in the FE layer due to an externally applied voltage is given by $|\vec{E}_{a,FE}| = \frac{V_a}{t_{FE}+\frac{\epsilon_{FE}}{\epsilon_{IL}}t_{IL}}$ as derived in Equation ES4 in the Supplementary Information. As such, one might

expect $V_{sw}$ to increase linearly with a slope of $E_c \frac{\epsilon_{FE}}{\epsilon_{IL}}$ as $t_{IL}$ increases. This "standard expectation" is plotted in Figure 4 (a). However, we have observed experimentally that $V_{sw}$ scaling is strongly sublinear and even exhibits an outright decrease for thin ILs. To understand this result, it is necessary to consider the effects of the depolarization field induced by the IL. We will first demonstrate our counterintuitive observation through two methods of extracting $V_{sw}$ and then theoretically derive the conditions of observing $V_{sw}$ reduction.

As described in the previous section, FE switching is accompanied by a modulation in the conduction mechanism that leads to an abrupt change of slope in the device's HRS *I-V* characteristics. As such, we can extract the $V_{sw}$ for devices of various IL thicknesses by finding the maximum values for the derivative, $\frac{d \log I}{dV}$. The derivative plots for five identical devices are included in figure S5 in the Supplementary Information. The results are shown in Figure 4(a). We can see that $V_{sw}$ decreases up until an IL thickness of 2 nm before increasing again. To further confirm this effect, "positive up negative down" (PUND) measurements were conducted at 200 kHz across all IL thicknesses as seen Figure 4(b). The applied voltage ($V_a$) necessary to get a $2P_r$ of 6 µC/cm² was subsequently extracted and shown in Figure 4(c). Confirming the earlier observations, an IL thickness of 1 or 2 nm lowered the required $V_a$. However, an increase in IL thickness beyond 3 nm corresponds with a higher $V_a$ than that in the control sample. We note that non-volatile resistance switching occurs even for partial polarization of the AlScN ferroelectric which is a sufficient condition for device operation. Hence, we consider a small $2P_r$ of 6 µC/cm² as a basis to evaluate if our observations are deviating from conventional theory.

We explain these observations by analyzing the electrostatics of the FE-diode using a single-domain Preisach model[12] to describe the behavior of the ferroelectric layer. We make this simplifying assumption and neglect multidomain effects here because it allows an informative analytical solution and the tight $E_c$ distribution of AlScN means that its hysteresis behavior closely resembles a square P-E loop implied by the single-domain model. We will follow with a more nuanced examination of multidomain effects in the following section.

In the single-domain model, the condition for switching is $E_{FE} = E_c$, where $E_{FE}$ is the local electric field in the ferroelectric layer and $E_c$ is the coercive field. By the principle of superposition, we can further write $E_{FE} = E_a + E_{dep}$ where $E_a$ is the electric field due to the applied bias and $E_{dep}$ is the depolarization field. The electric fields and the polarization state of the ferroelectric are shown graphically in Figure 4 (d) without an applied bias and (e) with an applied bias. A convention is taken to have the +y direction to be positive. Solving the Poisson equation for the device yields the following expressions:

$$\vec{E}_a = \frac{-V_a}{t_{FE} + \frac{\epsilon_{FE}}{\epsilon_{IL}} t_{IL}} \tag{2}$$

$$\vec{E}_{dep} = -\left(\frac{P_{FE} - \sigma_s}{\epsilon_{FE}}\right) \tag{3}$$

Where $t_{FE}$ and $t_{IL}$ denotes the thickness of the FE and IL layers respectively, and $\sigma_s$ is the screening charge in metal electrode interfaces given by

$$\sigma_s = \frac{P_{FE}}{1 + \frac{\epsilon_{FE}}{\epsilon_{IL}} \frac{t_{IL}}{t_{FE}}} \tag{4}$$

Substituting into the expression $E_c = E_{FE} = E_a + E_{dep}$ gives us

$$E_c = |\vec{E}_{FE}| = \frac{V_{sw}}{t_{FE} + \frac{\epsilon_{FE}}{\epsilon_{IL}} t_{IL}} + \left(\frac{\sigma_s - P_{FE}}{\epsilon_{FE}}\right) \tag{5}$$

and rearranging for $V_{sw}$ yields

$$V_{sw} = \frac{P_{FE} t_{FE}}{\epsilon_{FE}} + \left(t_{FE} + \frac{\epsilon_{FE}}{\epsilon_{IL}} t_{IL}\right)\left(E_c - \frac{P_{FE}}{\epsilon_{FE}}\right) \tag{6}$$

A step-by-step, explanatory derivation of this expression is included in the Supplementary Information.

As we can see, the effect of changing $t_{FE}$ and $\varepsilon_{FE}$ on $V_{sw}$ is dependent on the sign of the term ($E_c$-$P_{FE}$/$\varepsilon_{FE}$). If this term is positive, then decreasing IL capacitance (increasing $t_{FE}$ or decreasing $\varepsilon_{FE}$) will raise $V_{sw}$. However, if the term is negative, then decreasing IL capacitance will cause $V_{sw}$ to decrease. As such, the condition for achieving IL-induced $V_{sw}$ reduction is

$$E_c < \frac{P_{FE}}{\epsilon_{FE}} \tag{7}$$

Intuitively, this result expresses the idea that a small depolarization field can aid the switching of the ferroelectric, and the condition for an outright decrease in external switching voltage is if the depolarization field of the freestanding ferroelectric (i.e. without any electrode screening) is higher than the its coercive field. From the above derivation, we can see that this effect is best demonstrated on materials with high $P_r$, low $\varepsilon_{FE}$, and square-like hysteresis. AlScN satisfies these requirements well, but this result could apply to other ferroelectric systems such as $Hf_xZr_yO$, though square-like hysteresis have yet to be observed in such systems.

Lastly, we note the deviation from this theoretical result at large IL thicknesses where an increase in $V_{sw}$ is observed in Figure 4(a) and Figure 4(c). Our model for IL-induced $V_{sw}$ decrease is modeled using a first-order, single domain ferroelectric model, which means it is only highly accurate in the perturbative regime. Increasing the magnitude of the depolarization field through thicker ILs introduces higher order effects like the increased suppression of $P_{FE}$ by the depolarization field and increased deviation from a square-shaped hysteresis loop as suggested by our multidomain simulation described later in Figure 5(c). Nevertheless, there is only a modest increase in $V_{sw}$ compared to the control sample even when the IL thickness reached a maximum of 5 nm.

This result enables the tuning of operation voltages for ferroelectric diodes and ferroelectric tunnel junctions using the MFIM structure, which could be of high relevance for the further scaling of these devices to operating at CMOS-compatible voltages.

**Retention and Multi-State programming in 20 nm AlScN FE-diodes**

While the insertion of an IL brings multiple advantages, a thicker IL induces a higher $E_{dep}$, which could lead to retention degradation. Nevertheless, the 20 nm $Al_{0.72}Sc_{0.28}N$/4 nm $AlO_x$ FE-diode shows stable retention with an ON/OFF of 737 over $5\times10^4$ s as shown in Figure 5(a). Due to the high $E_c$ value of $Al_{0.72}Sc_{0.28}N$, even if $E_{dep}$ is significant in the 20 nm $Al_{0.72}Sc_{0.28}N$/4 nm $AlO_x$ device, the ratio of $E_{dep}/E_c$ remains low, ensuring good retention[1].

The introduction of a thicker IL in FE-diodes also makes these devices suitable for multi-state operation. This is due to their large ON/OFF ratios and their lower switching slope of the *P-E* loop, as suggested by our self-consistent device simulation program that implements a multi-domain Preisach FE model. A description of the model is included in the methods section, and the link to the associated code is available in the Supporting Information. The simulated *P-E* loops are shown in Figure 5(b). We observe that IL insertion stabilizes low $P_r$ states of the ferroelectric and sharply reduces the switching slope, enabling multi-state operation over a larger voltage range. The successful demonstration of multi-state FE-diode operation is depicted in Figure 5(c). More importantly, it was confirmed that 32 states can be stably maintained with a significant ON/OFF ratio of 1175, even beyond 300 seconds Figure 5(d). This robust performance in terms of maintaining multiple operational states indicates the considerable potential of these FE-diodes in advanced computational architectures and in-memory compute systems.

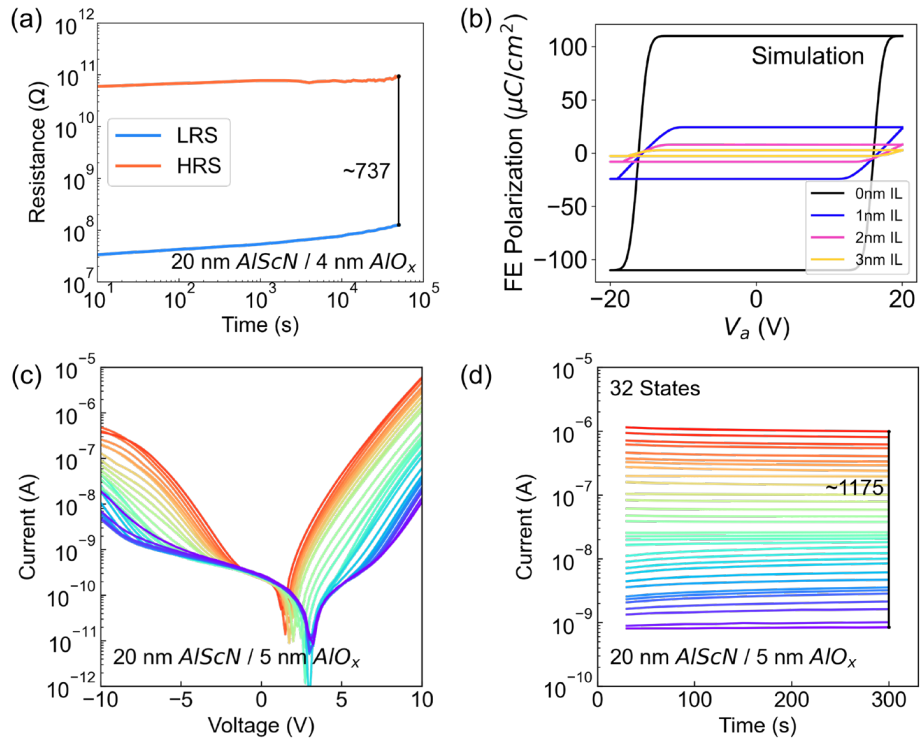

**Figure 5. Two-state retention, and multistate retention characteristics of 20 nm $Al_{0.72}Sc_{0.28}N$ FE-diode devices. (a)** LRS/HRS retention measurement of 20 nm $Al_{0.72}Sc_{0.28}N$ FE-diode with 4 nm $AlO_x$ IL up to 50,000 seconds. **(b)** Simulated IL thickness-dependent *P-E* hysteresis loop using a multidomain Preisach ferroelectric model. **(c)** 32-state *I-V* curves of the 20 nm $Al_{0.72}Sc_{0.28}N$ FE-diode with 5 nm $AlO_x$ IL FE-diode obtained after gradual switching by applying 0.25 V stepwise voltage amplitude. **(d)** 32 multi-states retention of FE-diode up to 300 s. The measurement scheme for the multi-state I-V curves is shown in the Supplementary Information Figure S4.

### FE-diodes based on 10 and 5 nm $Al_{0.72}Sc_{0.28}N/Al_{0.64}Sc_{0.36}N$

To achieve lower energy and lower voltage operation of the FE-diode, it is essential to further decrease the $V_{sw}$. To this end, we reduce the thickness of $Al_{0.72}Sc_{0.28}N$ to 10 and 5 nm as well as increase the Sc alloying to $Al_{0.64}Sc_{0.36}N$. It has been shown that increasing the Sc alloying decreases the $E_c$ of AlScN. Therefore, an increase of Sc% from 28% to 36% leads to lower operating voltages for our FE-diodes. In comparable devices with 10 nm thick AlScN films, the write voltage was lowered from 11 V (28% Sc) to 8 V (36% Sc).[13] Figure 6 (a-f) presents representative *I-V* curves of the 10 and 5 nm $Al_{0.72}Sc_{0.28}N/Al_{0.64}Sc_{0.36}N$ FE-diodes with Ti top electrodes as a function of the IL thickness. *I-V* curves of three devices of the same type are shown to confirm that device-to-device variation remains minimal even as the thickness of AlScN is reduced. Remarkably, the *I-V* curve profiles from three different devices are highly similar, highlighting reliable device uniformity. Both the ON/OFF and rectification ratios increase by 1-2 orders of magnitude when a thick $AlO_x$ IL was inserted for 10 and 5 nm $Al_{0.72}Sc_{0.28}N/Al_{0.64}Sc_{0.36}N$ FE-diodes (see Figure S2 for more details). Also, the nonlinearity of the device is important for practical circuit applications such as crossbar arrays, and all devices exhibit high nonlinearity (Supplementary Information Figure S6).

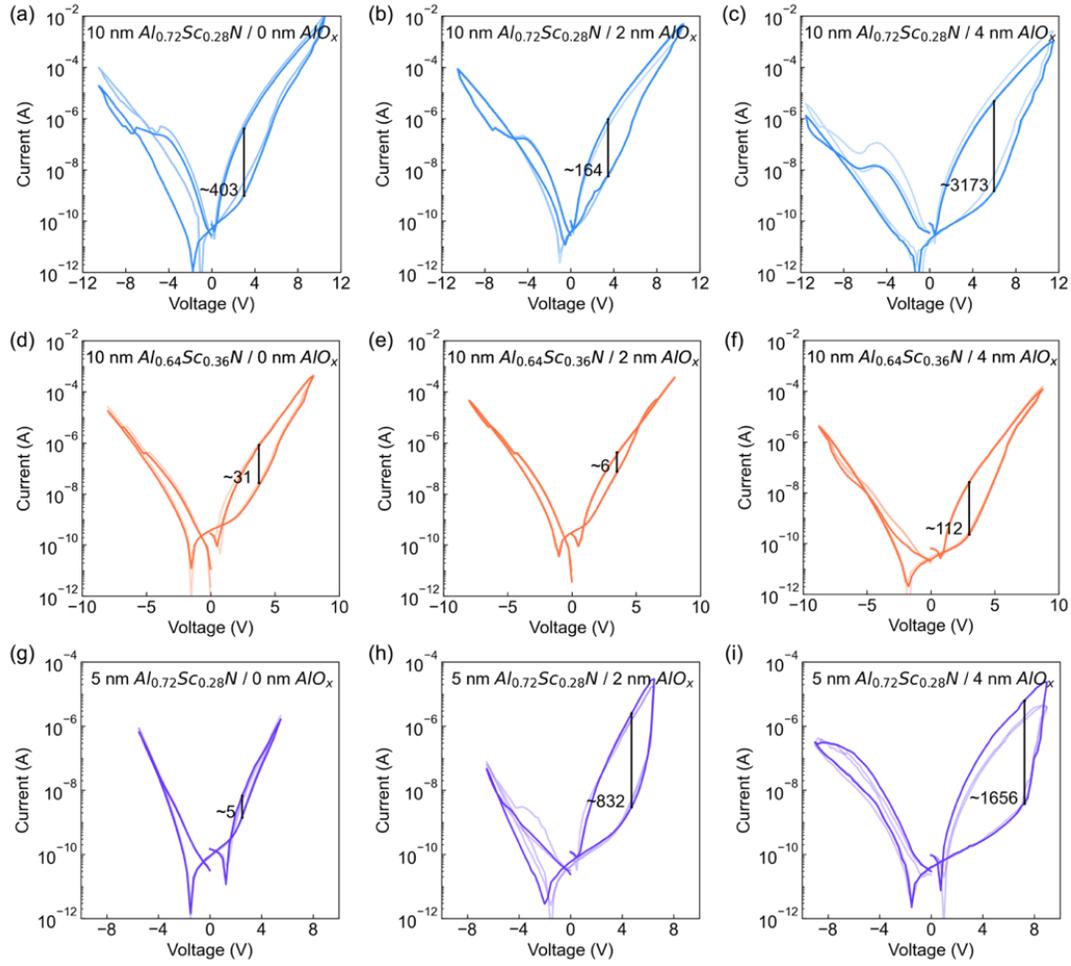

**Figure 6.** *I-V* characteristics of 10 and 5 nm $Al_{0.72}Sc_{0.28}N$/ $Al_{0.64}Sc_{0.36}N$ FE-diodes. 10 nm $Al_{0.72}Sc_{0.28}N$ FE-diode with $AlO_x$ IL thickness of **(a)** 0 nm, **(b)** 2 nm, and **(c)** 4 nm; 10 nm $Al_{0.64}Sc_{0.36}N$ FE-diode with $AlO_x$ IL thickness of **(d)** 0 nm, **(e)** 2 nm, and **(f)** 4 nm; 5 nm $Al_{0.72}Sc_{0.28}N$ FE-diode with $AlO_x$ thickness of **(g)** 0 nm, **(h)** 2 nm, and **(i)** 4 nm. Each plot includes *I-V* curves obtained from three different devices to confirm the low device-to-device variation. The insertion of a 4 nm IL into a 10 nm AlScN FE-diode results in a substantial enhancement in the ON/OFF ratio, increasing from 403 to 3173, as well as an improvement in the rectification ratio from 2992 to 5696 compared to the control sample (without IL). Similarly, for the 5 nm AlScN FE-diode, the introduction of the 2 nm IL leads to substantial improvements in the ON/OFF and rectification ratios, which increase from 5 to 832 and 14 to 2467 respectively. These results for scaled 10 and 5 nm devices follow the trend observed in 20 nm FE-diodes (Figure S2).

The physical model we developed for 20 nm $Al_{0.72}Sc_{0.28}N$ FE-diodes still provides an excellent fit to these scaled devices (see Figure S3 for additional details). Our model offers insight into why the magnitude of the increase is more pronounced in the 5 nm $Al_{0.72}Sc_{0.28}N$ compared to the 20 nm $Al_{0.72}Sc_{0.28}N$ case. The enhancements in the ON/OFF ratio for the 5 nm $Al_{0.72}Sc_{0.28}N$ FE-diodes exceed 330-fold, whereas the 20 nm $Al_{0.72}Sc_{0.28}N$ FE-diodes exhibit an increase of approximately 17 times. This disparity likely arises due to the change in the relative proportions of IL thickness to AlScN thickness. Specifically, in the 5 nm $Al_{0.72}Sc_{0.28}N$/4 nm IL configuration, the IL constitutes 80% of the combined thickness, while in the 20 nm $Al_{0.72}Sc_{0.28}N$/5 nm IL device, the IL only accounts for 25%. This greater proportion of the IL thickness leads to much larger IL-induced modulation in the potential barrier $\Phi$ as shown in Figure 3(b). The explicit dependence of barrier modulation on the ratio of IL/FE thickness was also derived in equation (1). This results in a more pronounced suppression of direct tunneling in the low voltage range. Consequently, a higher percentage of IL thickness relative to AlScN thickness appears beneficial for achieving a high ON/OFF ratio.

While a greater proportion of IL thickness to FE thickness can contribute to an enhanced ON/OFF ratio, theoretical predictions indicate that it also leads to an increased depolarization field as derived in equation (3). To confirm this, retention assessments are then conducted on the 10 and 5 nm $Al_{0.72}Sc_{0.28}N$ FE-diodes with ILs. The 10 nm $Al_{0.72}Sc_{0.28}N$/4 nm $AlO_x$ device shows a degradation in the ON/OFF ratio to 9 after $5\times10^4$ s (Figure 7(a)), and the 5 nm $Al_{0.72}Sc_{0.28}N$/2 nm $AlO_x$ sample lost its retention after $2.5\times10^3$ s (Figure 7(b)). This is because the proportion of the IL to the FE layer is substantial (40%), leading to a significant $E_{dep}$ that significantly reduces the retention.

Finally, we compare our AlScN FE-diodes based on 5 and 10 nm AlScN to other two-terminal ferroelectric NVM devices reported (Table 1). Our devices not only exhibit BEOL-compatibility at significantly reduced FE layer thicknesses, but also demonstrate competitive ON/OFF and rectification ratios. This suggests promising potential for their practical application in future deployment as dense, self-selective, low-power embedded memory in Si CMOS processors.

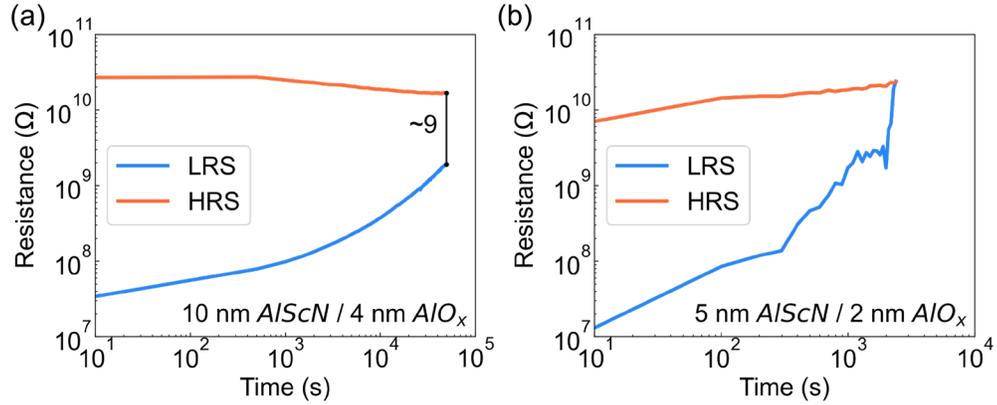

**Figure 7. Retention measurements of 5 and 10 nm Al$_{0.72}$Sc$_{0.28}$N FE-diodes. (a)** Retention measurement of the 10 nm AlScN/4 nm AlO$_x$ IL FE-diode up to $5 \times 10^4$ s and **(b)** the 5 nm AlScN/2 nm AlO$_x$ IL FE-diode up until the point of retention loss.

| Ferroelectric | CMOS BEOL compatibility | On-Off ratio | Thickness (nm) | Self-rectifying |
|---|---|---|---|---|
| BFO [5] | Low | 20,000 | 3 | Yes |
| HZO [6] | High | 10,000 | 10 | Yes |
| HZO [7] | Low (600 °C) | 300 | 12 | No |
| HZO [8] | Low (500 °C) | 100 | 9 | No |
| AlScN [9] | High | 50,000 | 20 | Yes |
| This work | High | 3,170 | 10 | Yes |
| This work | High | 1,650 | 5 | Yes |

**Table 1. Benchmarking table.** Comparison of the FE-diode in this work to other FE-diodes and ferroelectric tunneling junctions (FTJs) based on various ferroelectric materials.[2, 14-17] Maintaining CMOS BEOL-compatibility necessitates the ability to utilize existing CMOS processing equipment and to maintain a thermal budget under 400 °C. FE-diodes present self-rectifying *I-V* characteristics, thereby eliminating the need for a CMOS selector. In contrast, FTJs do not have this trait, potentially requiring a selector.

**CONCLUSION**

In summary, we report scaled FE-diode NVM based on AlScN with an AlO$_x$ IL showing high ON/OFF and rectification ratios in an all BEOL-compatible process. Furthermore, multi-state operation of the FE-diode is demonstrated, and retention is measured in all scaled devices. We present a comprehensive analysis on IL engineering with $AlO_x$ as an example, proving and emphasizing the need for higher quality and higher-k IL materials to improve retention, which constitute the most pressing challenges for further translation of AlScN FE-diodes for embedded memory applications.

## METHODS

**Al/Al$_{0.72}$Sc$_{0.28}$N/Al deposition.**

Aluminum (46 nm)/Al$_{0.72}$Sc$_{0.28}$N/Aluminum (46 nm) films were deposited onto 100 mm c-axis oriented sapphire substrates with an inclination of 0.2° +/- 0.1° to the M-plane. This was done using a PVD technique, specifically using the Evatec CLUSTERLINEVR 200 II system, with a distance of 88.5 mm between the target and substrate without breaking vacuum to avoid oxidation of Al$_{0.72}$Sc$_{0.28}$N. The <111> Al was deposited at 150 °C using an Al target power density of 12.7 W/cm$^2$ and an Ar flow rate of 20 sccm, leading to a process pressure of $1.1 \times 10^{-3}$ mbar. This resulted in the Al being deposited at a rate of 1.3 nm/s. The Al$_{0.72}$Sc$_{0.28}$N was deposited over the Al bottom layer by co-sputtering. This process was carried out at 350 °C, with Al and Sc target power densities of 11.1W/cm$^2$ and 7.07W/cm$^2$, respectively. The N$_2$ flow was set at 20 sccm, resulting in a process pressure ranging between $8 \times 10^{-4}$ and $8.5 \times 10^{-4}$ mbar. The thickness of Al$_{0.72}$Sc$_{0.28}$N was controlled by adjusting the deposition duration, with the deposition rate set at 0.25 nm/s. The 10 nm Al$_{0.64}$Sc$_{0.36}$N was deposited by altering the Sc cathode power density during co-sputtering to Al 900W; Sc 700W and increasing the N$_2$ flow during deposition to 30 sccm.

**Device fabrication.**

In the layered structure of Al/Al$_{0.72}$Sc$_{0.28}$N/Al films, the top Al layer was stripped off by immersing the sample in a 1% Hydrofluoric (HF) solution for 80 seconds. Subsequently, an Atomic Layer Deposition (ALD) process was carried out to deposit an Al$_2$O$_3$ layer on Al$_{0.72}$Sc$_{0.28}$N, utilizing the Cambridge Nanotech system from the USA. In this process, Trimethylaluminum (TMA) was employed as a metal-organic precursor, with water vapor introduced in each cycle. The top electrode patterns were intricately designed using electron-beam lithography. This was followed by the metal deposition process executed using electron-beam evaporation.

**Characterizations and device modeling/simulations of the FE-diodes.**

Transmission electron microscopy was performed on samples prepared using the focused ion beam and lift-out method. TEM was performed with a JEOL F200 instrument. The electrical measurements were performed in ambient air at room temperature using a Lakeshore probe station and a Keithley 4200A semiconductor analyzer.

The physical modeling of the device is based on the Poole-Frenkel emission and thermionic emission models with following two expressions for current density respectively:

$$J_{P-F} = AE \exp\left(-\frac{q\left(\Phi - \sqrt{qE/\pi\epsilon}\right)}{kT}\right) \quad (6)$$

$$J_{TI} = A^*T^2 \exp\left(-\frac{q\left(\Phi - \sqrt{qE/4\pi\epsilon}\right)}{kT}\right) \quad (7)$$

The multidomain simulation is based on the Preisach model and implemented through a custom Python program. The c/a ratio and associated $E_c$ and $P_s$ values of each domain are sampled from a Gaussian approximation to the Boltzmann distribution calculated by a previous report.[9] Closed-form solutions of the Poisson equation for the FE-diode are used in the simulation, and the average polarization state of the ferroelectric layer is iteratively updated until convergence at each step of the applied voltage.

**ASSOCIATED CONTENT**

Supporting information with Figure S1-S3 showing additional STEM images with composition mapping, extracted ON/OFF and rectification ratios of various FE-diode devices as well as additional I-V characteristics with theoretical fits to them.

# AUTHOR INFORMATION

## Author Contributions

K.-H. Kim and Z. Han contributed equally to this work.

D.J., R.H.O., K-H.K., and Z. H. conceived the idea of CMOS BEOL-compatible AlScN FE-diodes. K.-H. K. conceived the experiments and performed device fabrication. K.-H. K. and Z. H. wrote the manuscript with input from all authors. K.-H.K. and Z.H. performed the electrical measurements of the devices. Z. H. performed the physical device modeling and the simulation of ferroelectric properties. J.Z. performed sputtering to prepare the 5 nm, 10 nm, and 20 nm $Al_{0.72}Sc_{0.28}N$ substrates under the supervision of R.H.O. Y.Z. performed sputtering to prepare the 10 nm $Al_{0.64}Sc_{0.36}N$ films under the supervision of R.H.O. P.M. performed the electron microscopy and related characterizations. E.A.S. supervised the microscopy efforts.

## Notes

D.J., R.H.O., K-H.K., Z. H., and Y.Z. have filed a provisional patent based on this work. The authors declare no other competing financial interests.

# ACKNOWLEDGEMENT

D.J. and K-H.K. acknowledge primary support from the Air Force Office of Scientific Research (AFOSR) GHz-THz program FA9550-23-1-0391. R.H.O., Y. Z., J.Z. and P.M. acknowledge support from the Army/ARL via the Collaborative for Hierarchical Agile and Responsive Materials (CHARM) under cooperative agreement W911NF-19-2-0119. This work was carried out in part at the University of Pennsylvania Singh Center for Nanotechnology, which is supported by the NSF National Nanotechnology Coordinated Infrastructure Program under grant NNCI-1542153. The authors gratefully acknowledge use of facilities and instrumentation supported by NSF through the University of Pennsylvania Materials Research Science and Engineering Center (MRSEC) (DMR-1720530). Z.H. acknowledges funding support from the Vagelos Integrated Program in Energy Research (VIPER) at Penn.

Supporting information

# Multi-State, Ultra-thin, BEOL-Compatible AlScN Ferroelectric Diodes

Kwan-Ho Kim,[1,†] Zirun Han,[1,2,†] Yinuo Zhang,[1,3] Pariasadat Musavigharavi,[3] Jeffrey Zheng,[3] Dhiren K. Pradhan,[1] Eric A. Stach,[3] Roy H. Olsson III,[1] and Deep Jariwala[1,*]

[1]Department of Electrical and Systems Engineering, University of Pennsylvania, Philadelphia, Pennsylvania 19104, United States

[2]Department of Physics and Astronomy, University of Pennsylvania, Philadelphia, Pennsylvania 19104, USA

[3]Department of Materials Science and Engineering, University of Pennsylvania, Philadelphia, PA, USA

[†]These authors equally contributed to this work.

[*]Author to whom correspondence should be addressed: dmj@seas.upenn.edu

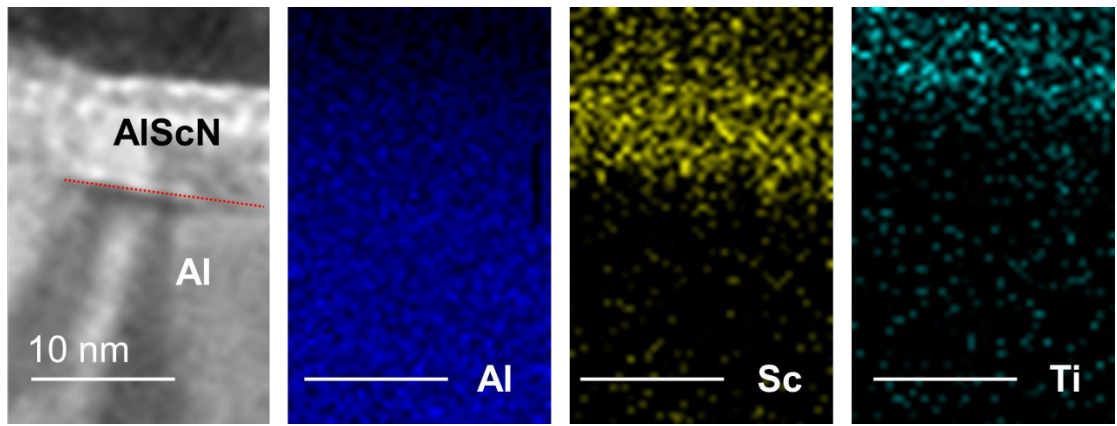

**Figure S1. STEM and EDX image of AlScN FE-diode.** STEM Spectrum image and corresponding energy-dispersive X-ray spectroscopy maps of the device.

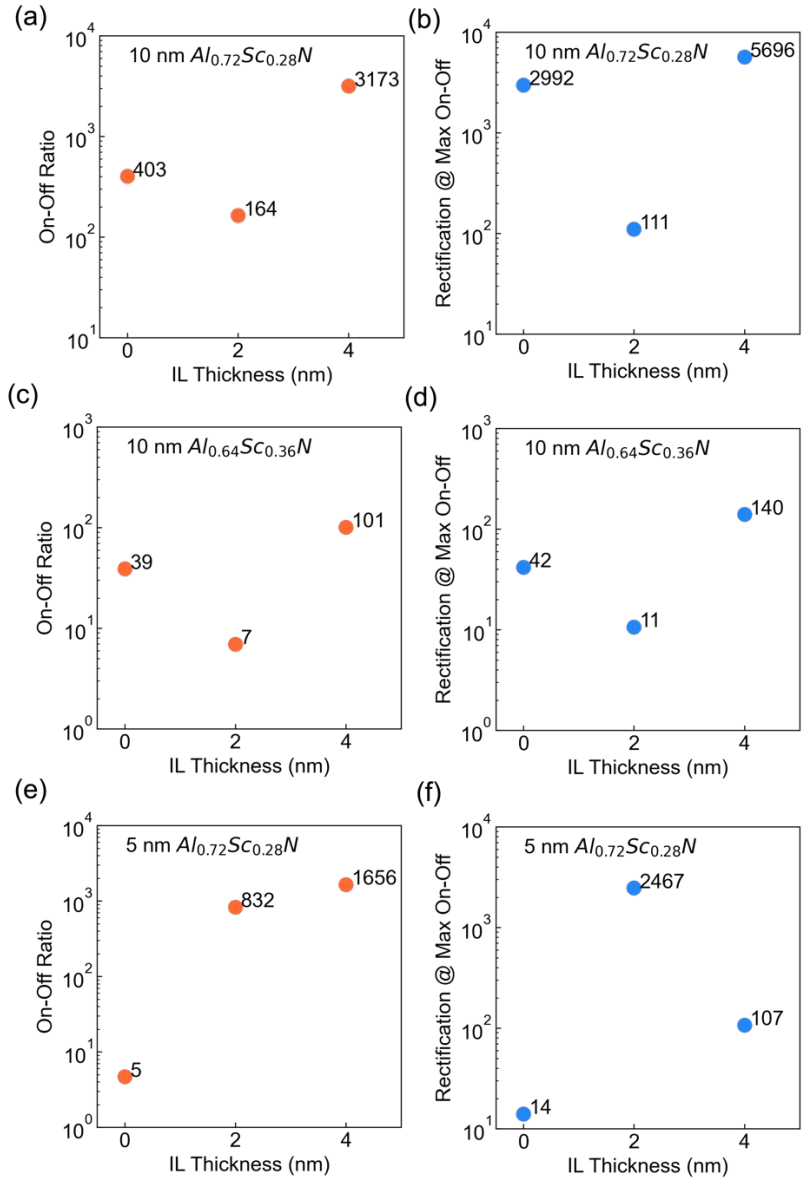

**Figure S2. (a-d)** Extracted On-Off ratio, and rectification ratio at the point of maximum On-Off ratio from *I-V* curves of FE-diodes with **(a, b)** 10 nm $Al_{0.72}Sc_{0.28}N$, **(c, d)** 10 nm $Al_{0.64}Sc_{0.36}N$, and **(e, f)** 5 nm $Al_{0.72}Sc_{0.28}N$ with various $AlO_x$ IL thicknesses (0-4 nm).

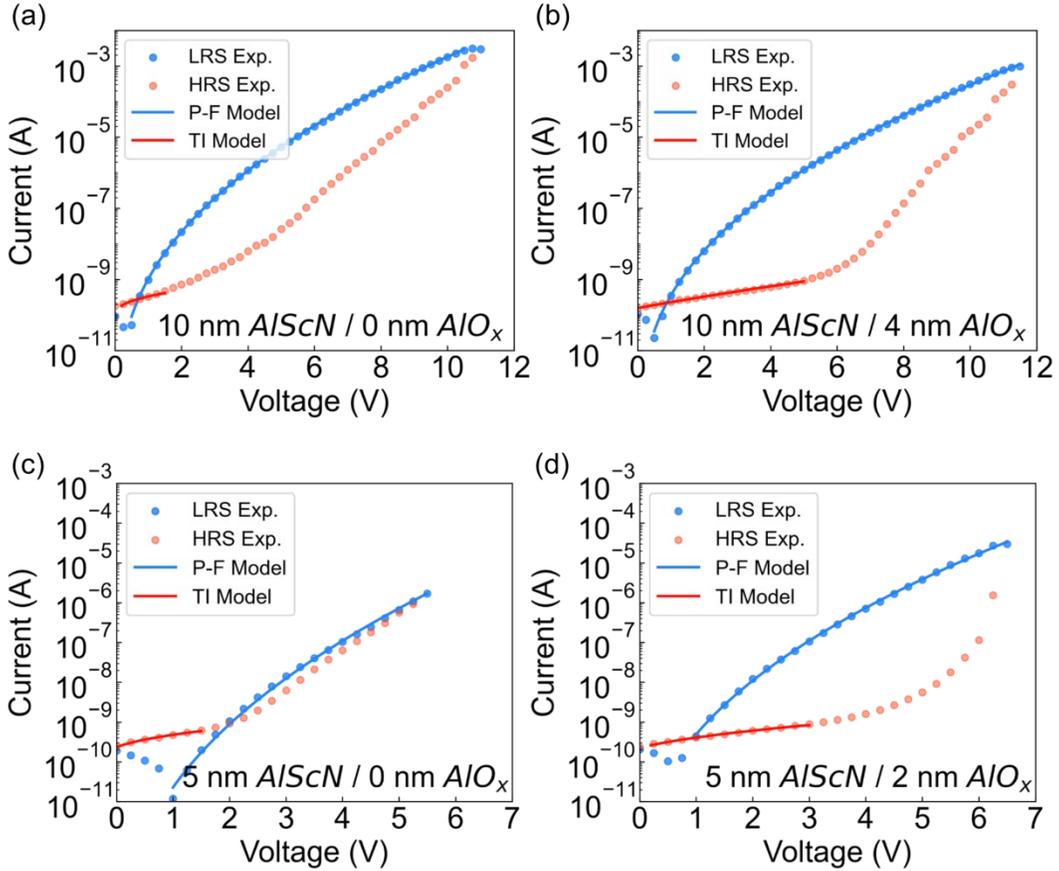

**Figure S3. Fitting of the Poole-Frenkel (P-F) and thermionic emission (TI) models** to **(a)** 10 nm AlScN/no IL, **(b)** 10 nm AlScN/2 nm AlO$_x$ IL, **(c)** 5 nm AlScN/no IL, and **(d)** 5 nm AlScN/4 nm AlO$_x$ IL FE-diodes. The dotted lines are experimental results, and solids lines are fitting models. Upon the introduction of the IL, the relatively flat region of HRS described by the TI model is extended, resulting in an enhanced On-Off ratio.

**Electrostatics of the FE-Diode and a Full Derivation for the Switching Voltage Reduction Model**

*Electrostatics of the FE-Diode*

We begin by considering the electric fields in the FE-diode caused by the polarization of the ferroelectric $P_{FE}$. Because the device diameter is much greater than its thickness, we can effectively approximate each layer of the device as infinite slabs. As such, $P_{FE}$ can be equivalently represented as a bound sheet charge density $\sigma_{FE} = \vec{P}_{FE} \cdot \hat{n} = P_{FE}$ at the terminating interfaces of the ferroelectric. We know that the electric field within the metal electrodes must be zero, so there must be a charge density $\sigma_s$ at the electrode interface that screens the ferroelectric polarization charges. At this point, we apply Gauss's law to derive the local electric fields in the IL and FE layers of the device:

$$\vec{E}_{P,FE} = \vec{E}_{dep} = -\left(\frac{P_{FE} - \sigma_s}{\epsilon_{FE}}\right) \#(ES1)$$

$$\vec{E}_{P,IL} = \frac{\sigma_s}{\epsilon_{IL}} \#(ES2)$$

Note we are aliasing $E_{P,FE}$ as $E_{dep}$, the depolarization field. We observe that the direction of this field must always oppose the direction of $P_{FE}$ because $\sigma_s \leq P_{FE}$.

We can then apply the short circuit condition $V(L) - V(0) = 0$ to obtain an expression for $\sigma_s$. Here, we assume the electric field is piecewise constant in the IL and FE and that there is zero trapped charge:

$$V(L) - V(0) = \left(\frac{\sigma_s - P_{FE}}{\epsilon_{FE}}\right) t_{FE} + \frac{\sigma_s}{\epsilon_{IL}} t_{IL} = 0$$

$$\Rightarrow \sigma_s = \frac{P_{FE}}{1 + \frac{\epsilon_{FE} t_{IL}}{\epsilon_{IL} t_{FE}}} \#(ES3)$$

Here, $t_{FE}$ and $t_{IL}$ are the thicknesses of the FE and IL layers respectively. We now proceed to calculate the electric fields in the device resulting from an externally applied voltage $V_a$. Here, we need to apply the continuity condition $\epsilon_{FE}\vec{E}_{FE} = \epsilon_{IL}\vec{E}_{IL}$ and use the fact that the total potential drop across the entire device is $V_a$. This yields

$$\vec{E}_{a,FE} = \frac{-V_a}{t_{FE} + \frac{\epsilon_{FE}}{\epsilon_{IL}} t_{IL}} \#(ES4)$$

$$\vec{E}_{a,IL} = \frac{-V_a}{t_{IL} + \frac{\epsilon_{IL}}{\epsilon_{FE}} t_{FE}} \#(ES5)$$

From the principle of superposition, the overall electric field is $\vec{E} = \vec{E}_P + \vec{E}_a$. At this point, we have a complete description of the electrostatics of the device.

*Switching Voltage Reduction Model*

At this point, we consider a first-order model for FE switching described by the single-domain Preisach model. In this model, the ferroelectric has a square-shaped hysteresis loop where switching occurs when local electric field in the ferroelectric $E_{FE} = \pm E_c$, the coercive field. This occurs when a sufficiently high external voltage $V_a$ is applied. We will call this the switching voltage $V_{sw}$. Combining equations ES1 and ES4, we have the condition for ferroelectric switching as

$$E_c = |\vec{E}_{FE}| = \frac{V_{sw}}{t_{FE} + \frac{\epsilon_{FE}}{\epsilon_{IL}}t_{IL}} + \left(\frac{\sigma_s - P_{FE}}{\epsilon_{FE}}\right)$$

Substituting for $\sigma_s$ as in ES3 and solving for $V_{sw}$ yields

$$V_{sw} = \frac{P_{FE}t_{FE}}{\epsilon_{FE}} + \left(t_{FE} + \frac{\epsilon_{FE}}{\epsilon_{IL}}t_{IL}\right)\left(E_c - \frac{P_{FE}}{\epsilon_{FE}}\right) \#(\text{ES6})$$

The implications of this expression are discussed in the main manuscript.

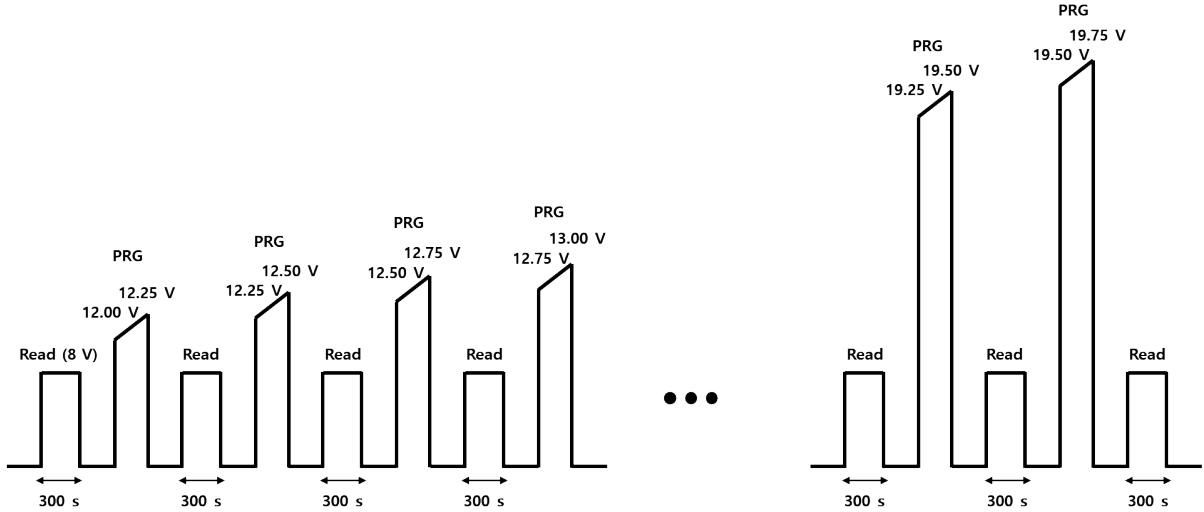

**Figure S4. Multistate measurement scheme.** During the programming (PRG) phase, two direct current (DC) voltages were subsequently applied, increasing in a step of 0.25 V from an initial value of 12.00 V up to a final value of 19.75 V. To be more specific, the first PRG aimed at partial switching ranged from 12.00 V to 12.25 V, followed by a second PRG phase that increased from 12.50 V to 12.75 V, continuing in similar increments until reaching the final voltage range of 19.50 V to 19.75 V. In the read phase, a periodic voltage of 8 V was applied at 30-second intervals across ten periods, culminating in a total duration of 300 second retention.

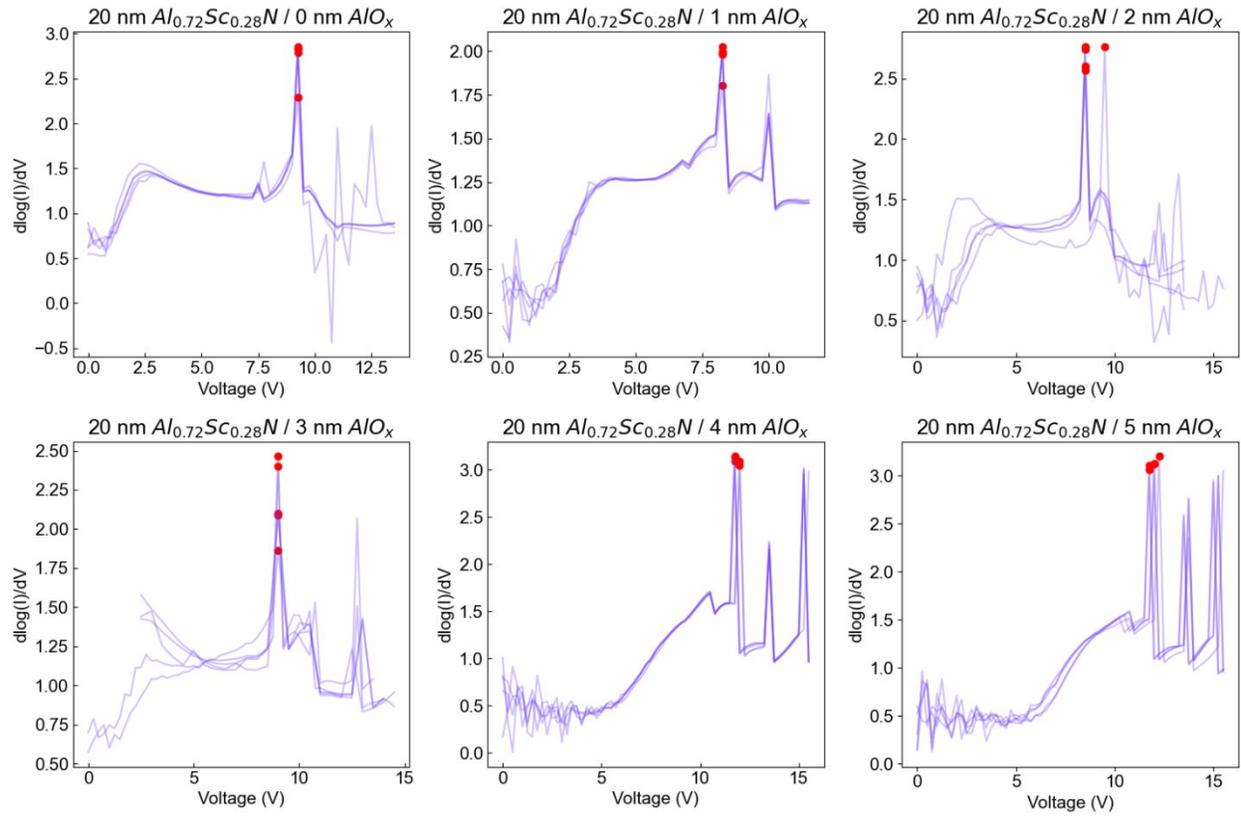

**Figure. S5. Derivatives of log *I-V* curves in Figure 2.**

| Device | $V_{read}$ | HRS Nonlinearity | LRS Nonlinearity |
|---|---|---|---|
| 20 nm $Al_{0.72}Sc_{0.28}N$/ 4 nm $AlO_x$ | 7.0V | 13 | 345 |
| 10 nm $Al_{0.72}Sc_{0.28}N$/ 4 nm $AlO_x$ | 6.5V | 21 | 706 |
| 5 nm $Al_{0.72}Sc_{0.28}N$/ 2 nm $AlO_x$ | 3.25V | 5 | 6038 |

**Figure. S6. Extracted nonlinearity of selected devices using the $V_{read}/3$ scheme.** $V_{read}$ is selected to be the voltage that gives the largest On-Off ratio. Nonlinearity is defined as $I(V_{read})/I(V_{read}/3)$.

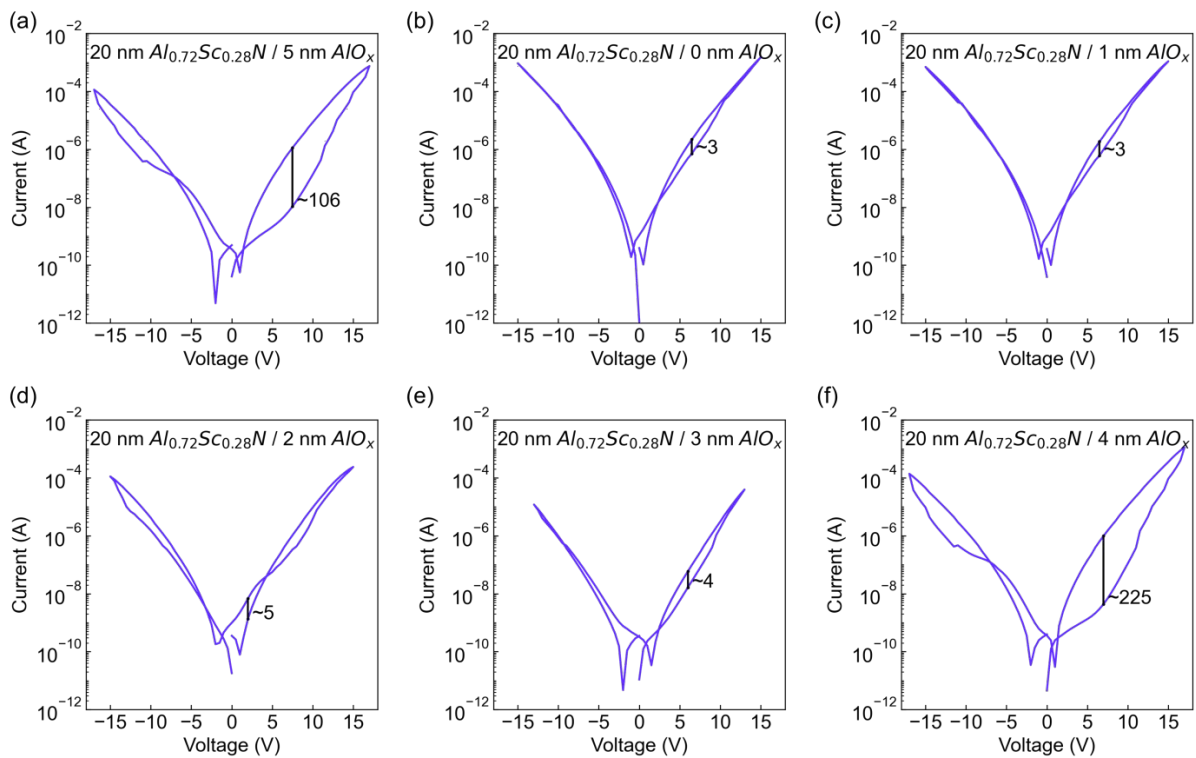

**Figure. S7. I-V curves of 20 nm Al$_{0.72}$Sc$_{0.28}$N ferroelectric diodes with Cr top contact.** I-V curves of 20 nm Al$_{0/72}$Sc$_{0.28}$N ferroelectric diodes with an AlO$_x$ thickness of **(a)** 0 nm, **(b)** 1 nm, **(c)** 2 nm, **(d)** 3 nm, **(e)** 4nm, and **(f)** 5 nm are shown.

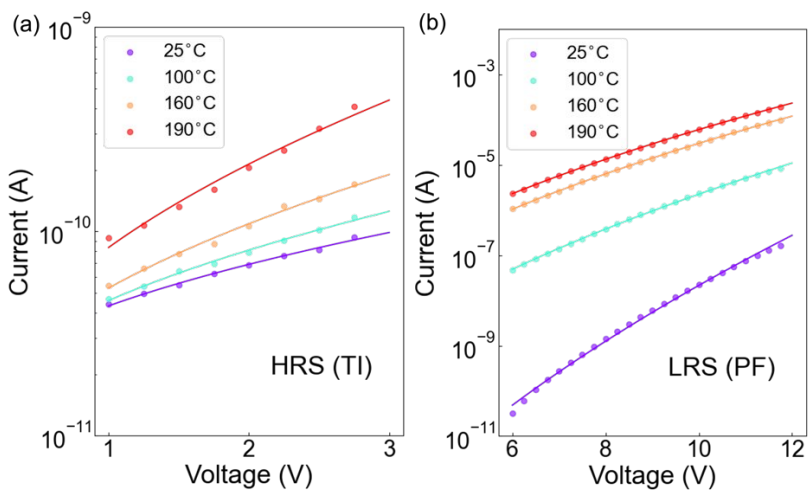

**Figure. S8. Temperature-dependent I-V data of the 20 nm Al$_{0.72}$Sc$_{0.28}$N / 5 nm AlO$_x$ device.** The experimental data is fitted to **(a)** the TI model in HRS and **(b)** the PF model in LRS.

**Python Code for the Multidomain Preisach Model-based Simulation**

The code used for the multidomain FE simulation in Figure 5(b) is available in the Github Repository linked here: https://github.com/danny1078/fed_multidomain_presaich/tree/main.